\documentclass[aps,prx,superscriptaddress,twocolumn,nofootinbib]{revtex4-1}
\usepackage{amsmath}
\usepackage{amsfonts}
\usepackage{graphicx}
\usepackage{color}
\usepackage{dsfont}
\usepackage{mathtools}
\usepackage[normalem]{ulem}
\usepackage{comment}
\usepackage{stackrel}
\usepackage{bm}
\usepackage[pdftex]{hyperref}
\hypersetup{colorlinks = true, urlcolor = black, linkcolor = black, citecolor = black}

\definecolor{myblue}{rgb}{.93, .93, 1}

\newcommand{\bsub}{\begin{subequations}}
	\newcommand{\esub}{\end{subequations}}

\begin{document}
	
	\title{Marginally localized edges of time-reversal symmetric topological superconductors}

\author{Yang-Zhi~Chou}\email{yzchou@umd.edu}
\affiliation{Condensed Matter Theory Center and Joint Quantum Institute, Department of Physics, University of Maryland, College Park, Maryland 20742, USA}

\author{Rahul~M.~Nandkishore}
\affiliation{Department of Physics and Center for Theory of Quantum
	Matter, University of Colorado Boulder, Boulder, Colorado 80309,
	USA}

\date{\today}

\begin{abstract}
We demonstrate that the one-dimensional helical Majorana edges of two-dimensional time-reversal symmetric topological superconductors (class DIII) can become gapless and insulating by a combination of random edge velocity and interaction. Such a gapless insulating edge breaks time-reversal symmetry inhomogeneously,
and the local symmetry broken regions can be regarded as static mass potentials or dynamical Ising spins. 
In both limits, we find that such gapless insulating Majorana edges are generically exponentially localized and trap Majorana zero modes.
Interestingly, for a \textit{statistically time-reversal symmetric} edge (symmetry is broken locally,  but the symmetry breaking order parameter is zero on average), the low-energy theory can be mapped to a Dyson model at zero energy, manifesting a diverging density of states and exhibiting marginal localization (i.e., a diverging localization length). 
Although the ballistic edge state transport is absent, the localized Majorana zero modes reflect the nontrivial topology in the bulk. Experimental signatures are also discussed.
\end{abstract}

\maketitle

\section{Introduction}

Topological insulators (TIs) and topological superconductors (TSCs) \cite{Hasan2010_RMP,Hasan2011,Schnyder2008,Qi2011_RMP,SenthilARCMP,Ludwig2015} possess nontrivial bulk winding numbers and host gapless conducting surfaces (or edges) that cannot be localized by disorder. Such delocalized surface states are anomalous and impossible to be realized in the bulk systems. In the presence of interaction, the clean TIs and TSCs can also allow for gapped surfaces \cite{SenthilARCMP} provided that the material boundaries either exhibit topological order or break the symmetry spontaneously. 
In particular, the avoidance of Anderson localization in the gapless surfaces is the hallmark of the topological protection and has opened up new possibilities to build novel topological electronic devices \cite{Alicea2012_review}.

The experimental characterizations of the TIs and TSCs mainly rely on the surface state properties \cite{Hasan2010_RMP,Hasan2011,Qi2011_RMP,Ando2013topological}. 
Therefore, it is important to understand the fate of TI/TSC boundaries with both interaction and disorder, even under somewhat extreme conditions \cite{Wu2006,Xu2006,Foster2012,Schmidt2012,Nandkishore2013,Foster2014,Kainaris2014,Xie2015,Santos2016,Chou2018,Chou2019_TIDrag,potirniche, Kagalovsky2018,Rachel2018,Chou2019,Kimchi2020}. For example, the delocalized surface states might be unstable to weak disorder and weak interaction \cite{Foster2012,Nandkishore2013,Foster2014}.
Surprisingly, gapless insulating symmetry broken edges \cite{Chou2018} or surfaces \cite{Chou2019,Kimchi2020} can even be realized due to the interplay of disorder and interaction, leading to localization. 
However, these localized boundaries of topological insulators are still anomalous and exhibit novel properties that can be viewed as a ``remnant of topology.'' This possibility of hosting localized boundaries in TIs and TSCs is important to the experimental characterization. In particular, the puzzling experimental results in InAs/GaSb \cite{Du2015,Li2015} may be explained by time-reversal broken localized edge states \cite{Chou2018} of two-dimensional (2D) TIs \cite{Kane2005_1,Kane2005_2,Bernevig2006}.

The occurrence of the localized edges \cite{Chou2018} in 2D time-reversal TIs can be understood by the inhomogeneous symmetry breaking mechanism. Since the single-particle backscattering is prohibited by time-reversal symmetry, the leading backscattering in the one-dimensional (1D) edge is due to the umklapp interaction. In the clean limit, the umklapp interaction is generically irrelevant unless the chemical potential is finely tuned. In the presence of disorder, umklapp interaction can become commensurate locally and enable local two-particle backscatterings that dynamically generate time-reversal breaking mass potentials. 
The symmetry broken edge realizes localization of $e/2$ charges, and the localization is nonmonotonic in the disorder strength \cite{Chou2018}. In addition, the localized edge preserves the $\pi$-flux anomaly of 2D TI \cite{Ran2008,Qi2008,Metlitski2019}, realizing an anomalous localization \cite{Kimchi2020}.
The above mentioned novel phenomena arise through the interplay of topology, disorder, and interaction. One cannot find an analog in the clean systems or in the noninteracting systems \cite{LPQO}.

The gapless insulating boundary is not particular to the 2D TIs. The same phenomenology can exist in three-dimensional (3D) TIs of class CII \cite{Ryu2012,Potter2017} and in the 3D topological crystalline insulators (TCIs) \cite{Ando2015}. In both the systems, the surface states form domains due to the inhomogeneous symmetry breaking (TIs) or disorder (TCIs). With a \textit{statistical symmetry} (symmetry is broken locally, but the symmetry breaking order parameter is zero on average), the domain walls percolate and the surface can be described by a disordered network model, realizing statistical topological insulators \cite{Fulga2014,Morimoto2015_AL}. The CII TI surface with a statistical particle-hole symmetry realizes a topological helical network \cite{Chou2019}, so does the TCI surface \cite{Ando2015}. In the presence of interaction, the helical network can be localized by the interaction within the domain wall, similar to the localization on 2D TI edges \cite{Chou2018}. Remarkably, the helical network can also be localized by the interaction at the network junctions and exhibit a \textit{clogged state} that is particular to the network model \cite{Chou2019}. 

The localized edges and surfaces of TIs can also be extended to the systems without charge $U(1)$ symmetry, e.g., TSCs.
In this work, we show that the helical Majorana edge state, as realized in 2D time-reversal TSC of class DIII \cite{Roy2008,Qi2009_TSC,Hasan2010_RMP,Qi2011_RMP}, can become gapless and insulating due to an interplay of disorder and interaction. 
Such a \textit{glassy} Majorana edge breaks time-reversal symmetry inhomogeneously and traps localized Majorana zero modes (MZMs). We show this via a mean field approximation and a mapping to random Ising model.
Intriguing, the glassy Majorana edge with a statistically time-reversal symmetry can be mapped to a Dyson model at zero energy, featuring diverging density of states and a \textit{marginal} localization (i.e., localization length diverges at zero energy). Our work establishes the first example of Dyson singularity on the boundary of a topological system.
Both the mechanism of the inhomogeneous symmetry breaking and the remnant signatures of topology are the primary focuses of this work.

\section{Model}

The edge state of a 2D time-reversal symmetric TSC (of class DIII) \cite{Roy2008,Qi2009_TSC,Hasan2010_RMP,Qi2011_RMP} is described by counter propagating Majorana fermions that form Kramers pairs. The clean noninteracting helical Majorana edge Hamiltonian is given by
\begin{align}\label{Eq:H_0}
	\hat{H}_0=v_0\int dx \left[\gamma_R\left(-i\partial_x\gamma_R\right)-\gamma_L\left(-i\partial_x\gamma_L\right)\right],
\end{align}
where $v_0$ is the velocity and $\gamma_R$ ($\gamma_L$) is the right (left) mover Majorana fermion field. The Majorana description is a consequence of the particle-hole symmetry. The edge state also satisfies time-reversal symmetry ($\mathcal{T}^2=-1$) with
the time-reversal operation implemented by $\mathcal{T}: \gamma_R\rightarrow \gamma_L$, $\gamma_L\rightarrow -\gamma_R$, and $i\rightarrow -i$. 

In the absence of interaction, the edge state is ballistic and avoids the Anderson localization as the elastic backscattering $\gamma_R\gamma_L$ is forbidden by the time-reversal symmetry. Different from the helical Luttinger liquid, the chemical potential is pinned to zero because of the particle-hole symmetry.
Therefore, the leading bilinear perturbation is in the form of ``velocity disorder.'' This velocity disorder is similar to the effect of random Rashba spin orbit coupling in a helical Luttinger liquid \cite{Xie2016}. 
The leading disorder effect can be characterized by
\begin{align}\label{Eq:H_dis}
	\hat{H}_{\text{dis}}=\int dx\, \delta v(x)\left[\gamma_R\left(-i\partial_x\gamma_R\right)-\gamma_L\left(-i\partial_x\gamma_L\right)\right],
\end{align}
where $\delta v(x)$ encodes the velocity fluctuation.
The local velocity is given by $v(x)=v_0+\delta v(x)$.
The velocity disorder alone cannot modify the ballistic nature of the helical Majorana edge state, i.e., avoidance of Anderson localization \footnote{This can be seen via a position-depedent rescaling such that $\tilde{\gamma}_R(x)\equiv \sqrt{v(x)/v_0}\gamma_R(x)$ and $\tilde{\gamma}_L(x)\equiv \sqrt{v(x)/v_0}\gamma_L(x)$. The Hamiltonian $\hat{H}_{0}+\hat{H}_{\text{dis}}$ becomes a ``clean'' Hamiltonian with a velocity $v_0$ in terms of $\tilde{\gamma}_R$ and $\tilde{\gamma}_L$. Therefore, the helical Majorana edge with velocity disorder remains ballistic.}. 
Microscopically, the velocity vanishes when the bulk gap closes. Therefore, we require that $v(x)>0$ since a well-defined bulk gap is considered.

The leading nontrivial interaction is at the quartic order and is analogous to a two-particle backscattering in helical Luttinger liquid \cite{Wu2006,Xu2006}. The interaction Hamiltonian is given by
\begin{align}
	\label{Eq:H_int}\hat{H}_{\text{int}}= U\int dx\, :\gamma_R(x)\gamma_R(x+\alpha)\gamma_L(x+\alpha)\gamma_L(x):,
\end{align}
where $U$ encodes the interaction strength, $\alpha$ is the ultraviolet length scale, and $:\mathcal{O}:$ denotes the normal ordering of an operator $\mathcal{O}$.
Formally, one needs to perform the following expansion $\gamma_{R/L}(x+\alpha)\approx \gamma_{R/L}(x)+\alpha \partial_x\gamma_{R/L}(x)$, and the interaction in Eq.~(\ref{Eq:H_int}) becomes to
\begin{align}
\hat{H}_{\text{int}}\approx U\alpha^2\int dx\, :\gamma_R(x)\left[\partial_x\gamma_R(x)\right]\left[\partial_x\gamma_L(x)\right]\gamma_L(x):.
\end{align}
This interaction is irrelevant in the renormalization group analysis, suggesting that the helical Majorana edge state is stable against infinitesimal $|U|$. However, a sufficiently large $|U|/v_0$ can open up a gap and break the time-reversal symmetry spontaneously \cite{Aasen2020}. 

The interacting disordered helical Majorana edge is described by $\hat{H}=\hat{H}_0+\hat{H}_{\text{dis}}+\hat{H}_{\text{int}}$ [Eqs.~(\ref{Eq:H_0}), (\ref{Eq:H_dis}), and (\ref{Eq:H_int})]. The main focus of this work is the time-reversal symmetry broken glassy Majorana edge which is realized by the interplay of strong disorder fluctuation in $v(x)$ and a strong interaction $|U|$. We note that a well-defined bulk gap is always assumed, i.e., disorder or interaction cannot induce a bulk phase transition.
In the next section, we discuss the mechanism of inhomogeneous symmetry breaking and analyze the edge state properties in detail.

\section{Mass generation due to symmetry breaking}

Due to the topological protection, the primary source of backscattering is the interaction in a helical Majorana edge. The clean interacting edge is characterized by $\hat{H}_0+\hat{H}_{\text{int}}$ [Eqs.~(\ref{Eq:H_0}) and (\ref{Eq:H_int})]. Infinitesimal $|U|$ cannot induce instability of the helical Majorana edge.
However, for a sufficiently large $|U|/v_0$, the edge state becomes gapped, and the time-reversal symmetry is broken spontaneously. The occurrence of a gap can be described by a time-reversal breaking mass term that is generated dynamically. 
At the level of mean field approximation, the interaction-driven mass term is given by
\begin{align}
	\hat{H}_M=2M\int dx \, \left(i\gamma_R\gamma_L\right),
\end{align}
where $M$ is the real-valued mass order parameter. Notice that the $i$ in the above expression is crucial to the hermitian property.

The four-fermion interaction in Eq.~(\ref{Eq:H_int}) can be viewed as a ``mass square'' interaction upto the normal ordering and the point splitting.
The value of $M$ can be derived by self-consistent calculations \cite{Aasen2020}, but the value of $M$ depends on the regularization scheme\footnote{In Ref.~\cite{Aasen2020}, the self-consistent calculations were performed with both a continuum model and a lattice model. The value of $M$ depends on which model is used.
In addition, there are certain qualitative difference: A first-order jump in mass was obtained in the continuum model, while a continuous phase transition was found in the lattice model approach.}.
These complications are not relevant for us. All that matters is that the mass order parameter becomes nonzero, $M\neq 0$, as long as $|U|/v_0>\Xi_c$, where $\Xi_c>0$ is the critical threshold of the dimensionless parameter. 
The sign of $M$ is arbitrary. We note that the mass order parameter is dynamical, i.e., it can fluctuate. To establish a static mean-field-like order, 
the edge length $L$ must be much larger than the coherence length $v_0/|M|$. Under the static mass assumption, the dispersion of the massive Majorana edge state is given by $E(k)=\pm\sqrt{(v_0k)^2+M^2}$, with an energy gap $2|M|$.

The main purpose of this section is to analyze the symmetry broken pattern in the helical Majorana edge with a velocity disorder. We first discuss the condition of realizing the local symmetry breaking. With the mean field approximation, we then map the symmetry broken Majorana edge to a Dirac fermion with a random mass. Such an edge state is gapless and insulating, and the mass domain walls trap MZMs.

\subsection{Local symmetry breaking}

\begin{figure}[t]
	\includegraphics[width=0.45\textwidth]{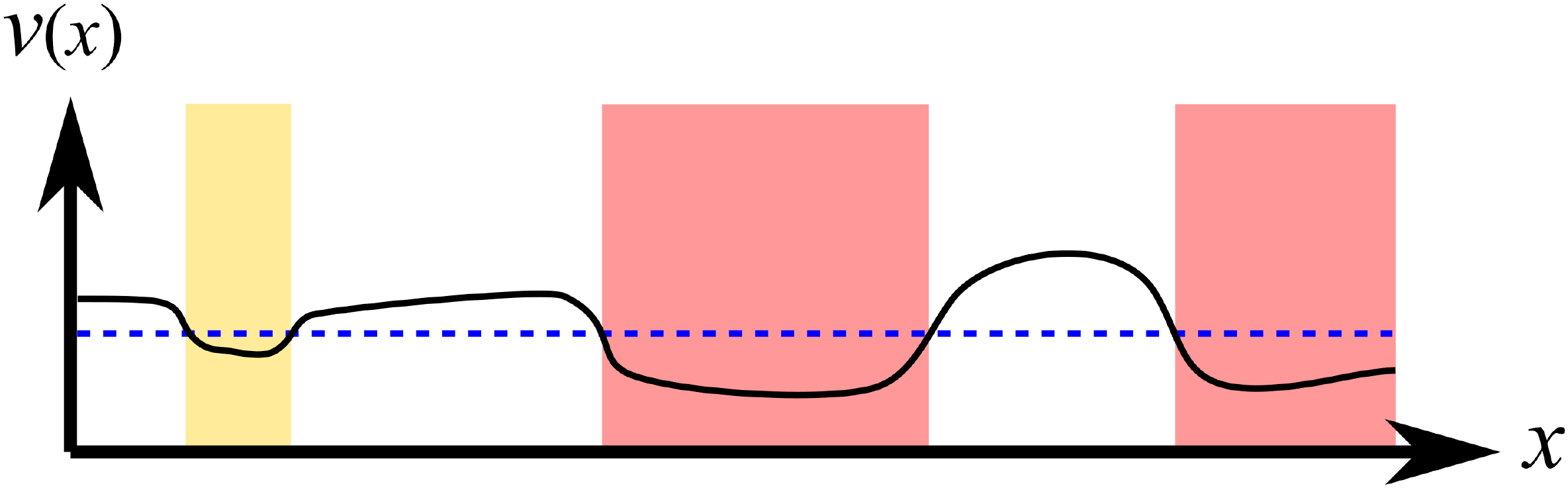}
	\caption{Velocity fluctuation and local symmetry breaking. The green curve indicates the velocity profile $v(x)$. When $v(x)<v_c\equiv|U|/\Xi_c$ (marked by the blue dashed line), local time-reversal symmetry breaking can take place. The mean-field-like masses are realized locally in the red regions. The mass order parameter is fluctuating inside the yellow region because the size of the yellow region is too small. No symmetry breaking happens in the white regions. See the main text for a detailed discussion.
	}
	\label{Fig:Inhomo_SSB}
\end{figure}

The disordered interacting edge state is described by $\hat{H}_0+\hat{H}_{\text{dis}}+\hat{H}_{\text{int}}$ [Eqs.~(\ref{Eq:H_0}), (\ref{Eq:H_dis}), and (\ref{Eq:H_int})]. As discussed previously, the condition of symmetry breaking is determined by the interaction-to-velocity ratio, $|U|/v_0$. In the disordered case, $v(x)=v_0+\delta v(x)$ is a positive-definite random function in space. Even with a uniform interaction strength $U$, this dimensionless parameter $|U|/v(x)$ fluctuates in space such that an inhomogeneous symmetry breaking might occur.

To gain intuitive understandings, we first assume that the velocity is slowly varying in space as illustrated in Fig.~\ref{Fig:Inhomo_SSB}. 
In such a limit, we can further approximate the system by multiple regions in which the velocity is essentially constant. 
For a given region label by $j$, we can define the region size and the velocity by $L_j$ and $v_j$ respectively. When $|U|/v_j>\Xi_c$, spontaneous symmetry breaking occurs, and a mass order parameter $M_j$ (determined by self-consistent calculations \cite{Aasen2020}) is generated dynamically. 
A mean-field-like order is realized when the region size is much larger than the correlation length, $L_j\gg v_j/M_j$. 
In such a situation, $M_j$ can be viewed as a quenched mass potential (red regions in Fig.~\ref{Fig:Inhomo_SSB}). 
The sign of $M_j$ is arbitrary as there is no explicit time-reversal breaking perturbation.
In the opposite limit ($L_j< v_j/M_j$), the mass order parameter is not frozen but fluctuating (yellow region in Fig.~\ref{Fig:Inhomo_SSB}). 
The fluctuating mass order parameter can be viewed as a fluctuating Ising spin, which we focus on in Sec.~\ref{Sec:Ising_spin}. The ground state is determined by the helical Majorana fermions scattering off the mass potentials and the fluctuating Ising spins.

Similar analysis also applies for a smooth disorder case, characterized by $\overline{\delta v(x)}=0$ and 
\begin{align}\label{Eq:dis_corr}
\overline{\delta v(x)\,\delta v(y)}= w^2\exp\left(-\frac{\left|x-y\right|^2}{2R^2}\right).
\end{align}
where $\overline{\mathcal{O}}$ denotes the disorder average of $\mathcal{O}$, $w>0$ encodes the strength of velocity fluctuation, and $R$ is the correlation length of the disorder potential. To fulfill the positive-definite condition of the velocity, we further require that $v(x)=v_0+\delta v(x)>0$.
The symmetry breaking pattern is controlled by both $w$ and $R$. We first define the critical velocity $v_c=|U|/\Xi_c$.
For $w>v_0-v_c$, we expect local symmetry breaking to prevail. For $w<v-v_c$, the local symmetry breaking happens only in certain rare regions. 
In addition, the local symmetry broken regions are of order size $R$. To develop mean-field-like massive regions, a sufficiently large $R$ or a sufficiently large $w$ is necessary. 
For $R\rightarrow 0$ (corresponding to the uncorrelated case),
the mean-field-like massive regions are rare, and most of the symmetry broken regions are fluctuating.

\subsection{Bound states}

The helical Majorana fermions can be reflected by the local symmetry broken regions. 
We first assume that the dynamically generated mass potential is completely frozen. The mass potential is position-dependent and described by
\begin{align}\label{Eq:H_M'}
	\hat{H}_M'= 2\int dx\, M(x) \left[i\gamma_R(x)\gamma_L(x)\right] ,
\end{align}
where $M(x)$ is the real-value mass parameter. 

Since the mass order parameter can be positive and negative, it is interesting to study the bound states sandwiched by masses with the same signs or with the opposite signs. We considered a helical massless Majorana edge of length $\mathcal{L}$ confined by two massive regions, $x<0$ and $x>\mathcal{L}$, with constant masses $M_1$ and $M_2$, respectively. We further assume a constant velocity $v$ in the massless region for simplicity.
We focus on the low-energy bound state spectrum, i.e. $|E|<|M_1|/2$ and $|E|<|M_2|/2$.
For $M_1M_2>0$, the momentum quantization corresponds to $e^{i2k\mathcal{L}}=-1$, and the bound state energy is given by $E=v|k|=\frac{v\pi}{2\mathcal{L}}(2n+1)$ where $n$ is an integer. The bound state can exist only if $E< \min(|M_1|/2,|M_2|/2)$ corresponding to a sufficiently large $\mathcal{L}$. Otherwise, two regions with $M_1$ and $M_2$ are merged into one massive region.
On the other hand, for $M_1M_2<0$, the momentum quantization corresponds to $e^{i2k\mathcal{L}}=1$, and the bound state energy is given by $E=\frac{v\pi}{\mathcal{L}}n$ with an integer $n$. The zero-energy state ($n=0$) is insensitive to the size $\mathcal{L}$. 
In the limit $\mathcal{L}\rightarrow 0$, the bound region is reduced to a zero-dimensional mass domain wall, and the zero-energy state becomes a MZM \cite{Aasen2020,Jones2019}. 
All the low-energy bound state energy levels do not depend on the $|M_1|$ and $|M_2|$ as long as the mass regions are sufficiently wide.

\subsection{Dirac fermion with a random mass}

Within the mean field approximation, the glassy Majorana edge can be described by
\begin{align}
	\nonumber\hat{H}_{\text{MF}}=&\int dx\, v(x) \left[\gamma_R\left(-i\partial_x\gamma_R\right)-\gamma_L\left(-i\partial_x\gamma_L\right)\right]\\
	\label{Eq:H_MF_random}&+2\int dx \left[i\gamma_R(x)\gamma_L(x)\right] M(x),
\end{align}
where $M(x)$ encodes the position-dependent real-valued mass potential.
To simplify the problem, we perform a rescaling such that $\tilde{\gamma}_R(x)\equiv \sqrt{v(x)/v_0}\gamma_R(x)$ and $\tilde{\gamma}_L(x)\equiv \sqrt{v(x)/v_0}\gamma_L(x)$.
This rescaling does not affect the localization property but eliminates the randomness in the velocity. Equation (\ref{Eq:H_MF_random}) is expressed by
\begin{align}
\nonumber\hat{H}_{\text{MF}}'=&\int dx\, v_0 \left[\tilde\gamma_R\left(-i\partial_x\tilde\gamma_R\right)-\tilde\gamma_L\left(-i\partial_x\tilde\gamma_L\right)\right]\\
	&+2\int dx \left[i\tilde\gamma_R(x)\tilde\gamma_L(x)\right] m(x),\\
\label{Eq:H_MF_random_resc}=&\int dx\, \Phi^T\left[-i\sigma^z\partial_x+m(x)\sigma^y\right]\Phi
\end{align}
where $m(x)=v_0M(x)/v(x)$ is the rescaled mass potential, $\Phi^T=[\tilde{\gamma}_R,\tilde{\gamma}_L]$ is a two component Majorana fermion field, and $\sigma^z$ ($\sigma^y$) is the $z$ ($y$) component Pauli matrix. Equation~(\ref{Eq:H_MF_random_resc}) describes
a 1D Dirac equation with a random mass (see \cite{McKenzie1996,Balents1997,Bocquet1999} and references therein), which is the low-energy theory of several models exhibiting Dyson singularity such as the random transverse field Ising model \cite{Fisher1992,Shankar1987,Fisher1995} and the random XY model \cite{Fisher1994}. 
This suggests that the glassy Majorana edge state can also realize the Dyson singularity, manifesting a diverging density of state and a marginal localization at $E=0$.
Note that we focus only on the zero-energy wavefunctions which are associated to our original Majorana theory.

To understand the localization properties, we investigate the eigenstate of Eq.~(\ref{Eq:H_MF_random_resc}).
For a given realization of disorder, the (two-fold degenerate) zero-energy wavefunctions can be obtained analytically \cite{Balents1997} as follows:
\begin{align}\label{Eq:wavefcn_E0}
\Phi_{0,\pm}(x)=\left[\begin{array}{c}
1\\
\pm 1
\end{array}
\right]e^{\mp\int_{\infty}^{x}dy\,m(y)/v_0},
\end{align}
where $\Phi_{0,\pm}(x)$ is the zero-energy wavefunction without normalization.
The logarithm of the $\Phi_{o,\pm}(x)$ can be viewed as a 1D random walk \cite{Balents1997} -- a particle ``moves'' $m(x)dx/v_0$ at ``time'' $x$ with a ``time step'' $dx$. The properties of the wavefunctions can be understood intuitively by this analogy to 1D random walk. We review the main results \cite{Balents1997} in the following.
The zero-energy states are exponentially localized generically as long as $m_0\equiv\overline{m(x)}\neq 0$ \cite{Balents1997}, and this is associated with spontaneous breaking of the protecting time reversal symmetry. The localization length is inversely proportional to $|m_0|$.
When $m_0= 0$ (``time-reversal symmetry on average''), a Dyson singularity develops at $E=0$, featuring a diverging density of states as well as a diverging localization length. We note that the zero-energy wavefunction given by Eq.~(\ref{Eq:wavefcn_E0}) is \textit{nodeless} and multifractal \cite{Balents1997}. Such a zero-energy wavefunction realizes a random singlet state \cite{Dasgupta1980}, demonstrating rarefied peaks with arbitrary peak-to-peak separations in a single wavefunction. 
The random singlet state gives rise to a stretched exponential behavior in the typical conductance, $G_{\text{typ}}\propto e^{-\mathcal{C}\sqrt{L}}$ \cite{Fisher1992,Fisher1994,Fisher1995,Balents1997,Steiner1999} where $L$ is the length of the edge and $\mathcal{C}$ is a constant depending on the fluctuation of the random mass potential. 
We note that this is different from the case in the localized TI edge \cite{Chou2018}, where exponential localization is expected generically. It corresponds instead to what was referred to in \cite{NandkishorePotter, Nandkishore2d} as ``marginal localization.''

The mapping to a quenched random mass Dirac equation is not fully justified, because going beyond the mean field approximation, the mass potential can fluctuate, and the results obtained based on static mass profile need to be scrutinized. In particular, the random singlet wavefunctions and the stretched exponential correlation for the statistical time-reversal symmetric glassy Majorana edges might not be valid anymore.
It is thus important to consider the fluctuations of the mass potential, which we turn to next.

\section{Effective random Ising Model}\label{Sec:Ising_spin}

\begin{figure}[t]
	\includegraphics[width=0.35\textwidth]{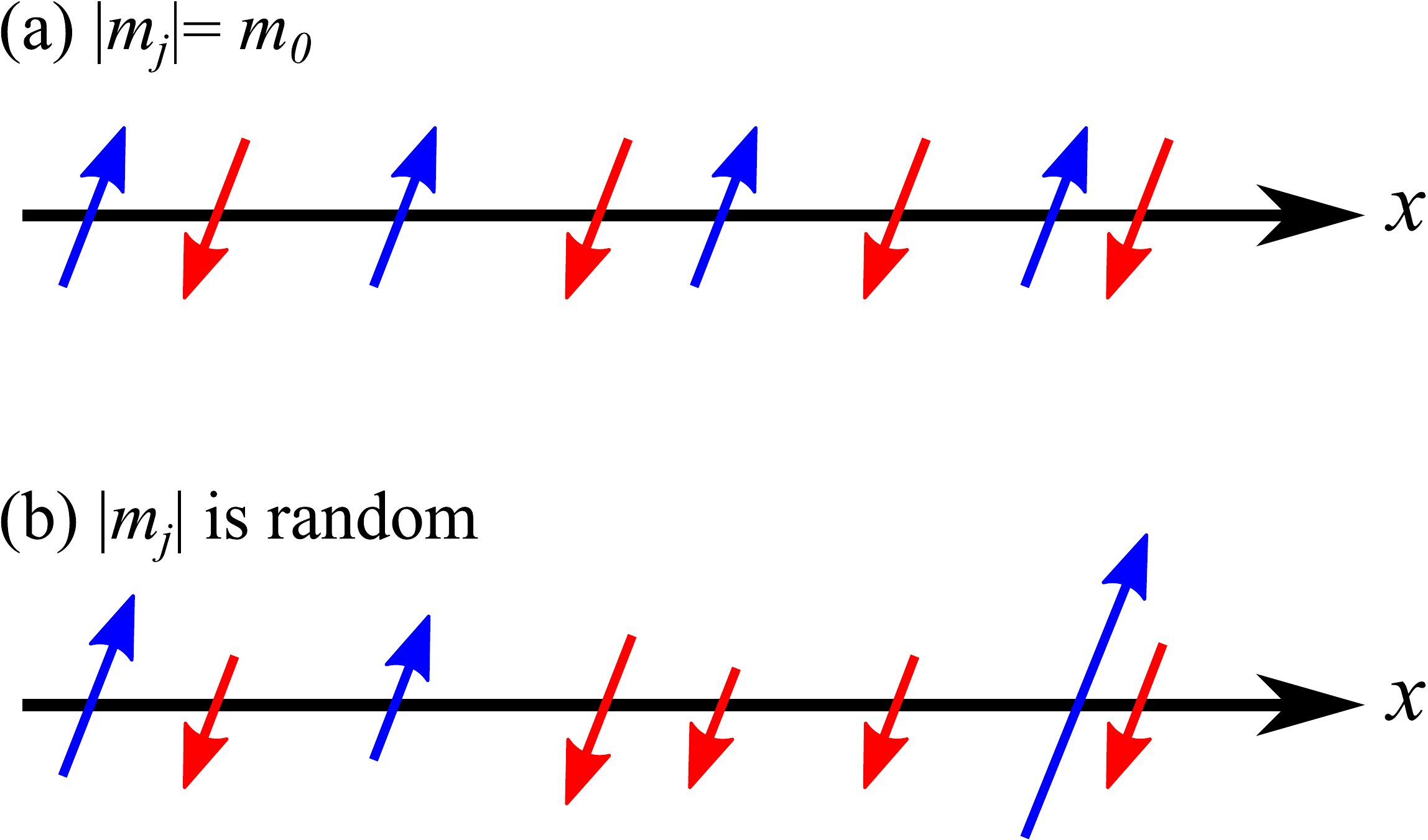}
	\caption{The spin configuration of the random $1/r$ interacting Ising model given by Eq.~(\ref{Eq:H_eff_N}). We use blue and red colors to mark the up spins and down spins respectively. (a) When $|m_j|=m_0$ for all $j$, the ground state is a Ne\'el order. (b) When $|m_j|$ is random, the ground state is more complicated. As illustrated, the rare huge spin (second from the right) screen the nearby spins. Local Ne\'el order is still expected away from such a rare region.
	}
	\label{Fig:Ising}
\end{figure}

When we move beyond the mean field, we allow for the fact that the mass order parameter can fluctuate. Fluctuations will be weak in the limit of large disorder correlation length $R$ [in Eq.~(\ref{Eq:dis_corr})], but conversely in the limit $R\rightarrow 0$ strong fluctuations are guaranteed. To study the fluctuation, we map the local mass potentials to dynamical Ising spins in the following manner. First of all, we assume there are $N$ symmetry broken impurity regions. These symmetry broken impurities are marked by a set of positions $\{x_1,x_2,\dots,x_N\}$ with the impurity mass order parameters $\{m_1,m_2,\dots,m_N\}$ respectively. We order the positions such that $x_1<x_2<..<x_N$.
The coupling between the Majorana fermion and the local mass is an Ising coupling [given by the second line of Eq.~(\ref{Eq:H_MF_random_resc})].
Treating the Ising coupling perturbatively, one can derive an effective RKKY interaction mediated by the Majorana fermions \cite{Eriksson2015}. For two impurities located at $x_1$ and $x_2$, the effective interaction is given by \cite{Eriksson2015}
\begin{align}
	\hat{H}_{\text{eff},I}^{(2)}=\frac{\left|m_1m_2\right|}{4\pi v_0|x_1-x_2|}\tau_1^z\tau_2^z,
\end{align} 
where $\tau_j^z=\pm 1$ is the Ising variable keeping track of the sign of $m_j$.
This effective spin-spin interaction is antiferromagnetic and decays with $1/r$. The absence of the oscillation is due to the particle-hole symmetry (i.e., $k_F=0$). For $N$ impurities, the interacting Ising Hamiltonian is as follows:
\begin{align}\label{Eq:H_eff_N}
	H_{\text{eff},I}=\sum_{j<j'}\frac{\left|m_jm_{j'}\right|}{4\pi v_0|x_j-x_{j'}|}\tau_{j}^z\tau_{j'}^z.
\end{align} 
When all the mass impurities have the same amplitude ($|m_j|=m_0$ for $j=1,\dots,N$),
the ground state is a Ne\'el state [Fig.~\ref{Fig:Ising}(a)]. In general, a huge mass impurity can ``screen'' spins beyond nearest neighbor, but the Ne\'el like order is still expected locally [Fig.~\ref{Fig:Ising}(b)]. We can also view this problem as a 1D Coulomb glass system \cite{Raikh1987} except that the charges of the particles can be arbitrary in our case. 

The $1/r$ spin-spin interaction is classical, i.e., the Ising spins cannot be flipped. 
Our goal is to explore the stability of the classical spin configuration in the presence of quantum mechanical fluctuation (spin flipping). Thus, we undertake a phenomenological approach rather than an explicit microscopic derivation.
To incorporate the fluctuations, we consider a phenomenological transverse field term as follows:
\begin{align}\label{Eq:H_eff_h}
	H_{\text{eff},h}=-\sum_jh_j\tau_j^x,
\end{align}
where $h_j$ encodes the degree of fluctuation of the impurity $j$ and $\tau^x_j$ is the $x$-component of the Pauli matrix flipping the $j$th spin. Importantly, the larger the $|m_j|$, the smaller the $|h_j|$. We assume that the sign of $h_j$ is arbitrary.

The Hamiltonian $H_{\text{eff},I}+H_{\text{eff},h}$ given by Eqs.~(\ref{Eq:H_eff_N}) and (\ref{Eq:H_eff_h}) describes an interacting $1/r$ antiferromagnetic Ising chain with a random transverse field. 
We argue that, since the short range Ising chain is stable to a weak transverse field (at zero temperature), 
and the long range coupling should only make the order more stable, accordingly the classical Ising configuration should be stable. 
To gain further insights, we view the transverse Ising model as a 1D $1/r$ hopping model with disorder, where $H_{\text{eff},I}$ is the $1/r$ hopping, and $H_{\text{eff},h}$ acts like a quenched disorder. The rate of finding a resonance is proportional to $\int dr/r\sim \log L$ ($L$ being the system size), which is logarithmic divergent at large distances. Therefore, a resonant hopping process is always guaranteed as long as the system is sufficiently large. 
This situation is the same as the the $1/r^3$ hopping problem in the three dimensions. Since the 3D $1/r^3$ hopping model is critically delocalized \cite{Levitov1990}, we expect that the 1D $1/r$ hopping model is also critically delocalized. 
However, the $1/r$ term in our model is not the hopping, but rather the Ising interaction. The results in Ref.~\cite{Levitov1990} suggest that $H_{\text{eff},h}$ is unlikely to destroy the ordered state set by $H_{\text{eff},I}$.
Therefore, we conclude that the ground state of the random Ising model is essentially \textit{classical} and can be approximated by a static spin configuration.

In the low-energy limit, the helical Majorana fermions scatter off the asymptotic static Ising spins and become insulating. We expect that the predictions based on the mean field approximation remain valid in this situation - exponential localization takes place for generic Ising spin configurations, and the Dyson physics (marginal localization) arises when $\sum_{j=1}^{N}m_j=0$.

\section{Discussion}

We study a dirty interacting helical Majorana edge state as realized on the boundary of 2D time-reversal TSC of class DIII. We show that the inhomogeneous time-reversal symmetry breaking can happen due to a combination of random velocity and strong interaction. The regions with time-reversal broken order can be treated as a static mass order (mean field approximation) or an Ising spin (fluctuating limit). In both limits, we expect that the edge state becomes gapless and insulating. Since the mass domain walls trap MZMs, and the localized edge can be thought as a localization of MZMs. We also point out that a stretched exponential transport behavior can occur for an edge with a statistically time-reversal symmetry, i.e. the conductance $G$ scales with systems size $L$ as $G \sim \exp(- C \sqrt{L}$), with this unusual behavior originating from a Dyson singularity in the density of states and localization length.  

It is interesting to compare the glassy Majorana edge to the localized state as realized on 2D TI edges \cite{Chou2018}. 
Under time-reversal symmetry breaking, the helical Majorana edge acquire one type of mass term, while there are two types of masses in the helical Luttinger liquid. 
This difference is intrinsic to the $U(1)$ charge conservation.
The domain walls of the Ising mass trap MZMs in the helical Majorana edges. On the other hand, the domain walls of each type of masses trap half charges (Jackiw-Rebbi soliton \cite{JackiwRebbi1976}) in the helical Luttinger liquid. 
The localization properties are also intriguing. The glassy Majorana edge is generically exponentially localized except that random singlet states can take place for a statistical time-reversal symmetry (i.e., the mass or Ising spin is averaged out to be zero). In a localized TI edge, nonmonotonic localization is expected, i.e., the localization length is the shortest for an intermediate disorder strength \cite{Chou2018}. Both the glassy Majorana edges and the localized TI edges are examples of the non-Fermi glass \cite{Parameswaran_NF_glass} - localized states not adiabatically related to Anderson localization.

Finally, we discuss possible experimental characterizations of the gapless insulating helical Majorana edges. Since the Majorana fermions do not carry a charge but carry heat, the thermal transport is necessary to detect the insulating behavior. In addition, the regions with time-reversal breaking orders and the MZMs trapped by the domain walls can be detected by \textit{in situ} measurements, such as a scanning tunneling microscope \cite{Nadj-Perge2014} (for MZMs) and SQUID \cite{Nowack2013,Spanton2014} (for local time-reversal breaking order).  
We also would like to point out that our theory can apply to the ``sewing of Majorana edges'' in the novel setup for detecting the Kitaev spin liquid electrically \cite{Aasen2020}. 
Our work establishes that the glassy Majorana edge states can arise from a 2D time-reversal symmetric TSC bulk. The unconventional localization is enabled by the interplay of topology, disorder, and interaction.

\section*{Acknowledgments} 

Y.-Z.C. thanks Zhentao Wang for useful discussions. We also thank Leo Radzihovsky for previous collaborations on the localized TI edges.
This work was supported by the Laboratory for Physical Sciences (Y.-Z.C.) and by JQI-NSF-PFC (supported by NSF
grant PHY-1607611, Y.-Z.C.). Research was sponsored by the Army Research Office (R.M.N.) and was accomplished under Grant Number W911NF-17-1-0482. The views and conclusions contained in this document are those of the authors and should not be interpreted as representing the official policies, either expressed or implied, of the Army Research Office or the U.S. Government. The U.S. Government is authorized to reproduce and distribute reprints for Government purposes notwithstanding any copyright
notation herein.

%%%\bibliography{TI_ref}

%%%%%

\end{document}